\documentclass[11pt]{article}

\usepackage[preprint]{acl}

\usepackage{times}
\usepackage{latexsym}

\usepackage[T1]{fontenc}

\usepackage[utf8]{inputenc}

\usepackage{microtype}

\usepackage{inconsolata}

\usepackage{graphicx}

\usepackage{booktabs}

\usepackage{enumitem}

\usepackage{amsmath}
\usepackage{amssymb}

\title{RSRank: Learning Relevance from Representational Shifts}

\author{
    \textbf{Archit Gupta}$^{1}$
    \quad
    \textbf{Sai Sundaresan}$^{1}$
    \quad
    \textbf{Debabrata Mahapatra}$^{1}$\thanks{Corresponding author: dmahapatra@adobe.com} \\\\
    $^{1}$Adobe Research, India \\\\
}
\begin{document}
\maketitle

\begin{abstract}
As enterprises deploy RAG\nobreakdash-based systems to provide grounded responses to user queries, reranking has become a critical component for the final filtering step that separates relevant from distracting or irrelevant documents. Existing rerankers often rely on heuristic thresholds to achieve optimal filtering. Moreover, for relevance scoring, state-of-the-art methods use a language model's logit signals, which are designed for next-token prediction, not for assessing relevance. To address these limitations, we identify a principled signal for relevance: the \textbf{representational shift (RS)} induced in a query's internal state when conditioned on a document. We observe that the alignment between (a) RS induced by a candidate document and (b) RS induced by an oracle document-set provides a robust indicator of relevance. Building on this insight, we introduce a lightweight training framework that learns projections mapping RS to calibrated relevance scores. Our training objectives naturally filter irrelevant content at a zero threshold, reducing dependence on heuristic tuning. Across diverse retrieval datasets, our method delivers gains over SOTA rerankers.
\end{abstract}

\section{Introduction}
\label{sec:introduction}

\subsection{Role of Rerankers in Enterprise Systems}

Information retrieval (IR) systems form foundational infrastructure for enterprises that serve users at scale. Search engines, knowledge bases, and AI assistants increasingly depend on their ability to identify relevant information from a large corpus for a user query. A fundamental tradeoff that governs the design of these systems is \emph{efficiency} vs. \emph{accuracy}. Traditional methods like BM25~\citep{bm252009robertson} and modern dense embedding approaches~\citep{colbert2020khattab, nogueira-etal-2020-document} are efficient, as documents are encoded once and indexed for fast lookup, but suffer from an information bottleneck, being unable to represent query-specific information when constrained to a single document representation~\citep{luan-etal-2021-sparse}.

The introduction of reranking through a two-stage retrieval pipeline~\citep{cascade2011wang} addresses this representational limitation. The first stage retrieves a broad candidate set using efficient methods (accepting some loss in precision), and the second stage applies more expensive models to \emph{rerank} these candidates. Rerankers jointly represent query and document tokens, capturing semantic relationships that independent encodings miss, improving retrieval quality~\citep{nogueira2020passagererankingbert}.
More recently, LLM-based rerankers have extended this paradigm, achieving SOTA results~\citep{qwen3embedding}. This two-stage architecture is now standard in enterprise systems~\citep{ecom2017liu, semantic2021microsoft}.

\subsection{Reranking in Retrieval Systems}

Retrieval systems~\citep{rag2020lewis} have widely adopted rerankers, with vector database platforms providing native reranker support~\citep{pinecone2024rerankers, weaviate2024cohere}, and cloud providers offering reranking through APIs~\citep{aws2024rerank, google2025ranking}. In these systems, retrieved documents enter the LM's context window, and irrelevant documents degrade response quality~\citep{liu-etal-2024-lost, wu2024howcolm}, increase latency, and raise API costs~\citep{pinecone_reranking_2024}. Accurate selection is therefore critical, and the tradeoff between accuracy and efficiency becomes even more consequential in this setting.

In practice, the standard approach is to retrieve a broad candidate set (e.g., top-100) via embedding search, rerank, and then select a subset for the LM's context. The selection step is typically performed via fixed top-$k$ selection or score thresholding.

\subsection{Limitations of Existing Rerankers}

Neither approach discussed above adequately addresses the efficiency--accuracy tradeoff. A fixed top-$k$ selection ignores that queries differ in the number of relevant documents, often selecting too many or too few documents per query. Determining a score cutoff is also difficult because rerankers are not calibrated for absolute relevance; optimal values vary across domains and even across queries. We quantify these inefficiencies empirically in Sec.~\ref{sec:motivation}. 

These limitations stem, in part, from how rerankers derive their signal. Current approaches rely on signals tuned for next-token prediction (internal states, attention maps, logits), rather than relevance assessment~\citep{qwen3embedding, chen-etal-2024-m3, chen2025attention}. The resulting scores are effective for \emph{ranking}---ordering documents by relevance---but poorly calibrated for \emph{selection}---deciding which documents are relevant.

\subsection{Toward a Calibrated Relevance Signal}

The limitations above motivate a search for a different relevance signal---one that is inherently calibrated rather than repurposed from the next-token prediction objective. 
We observe that relevance fundamentally concerns how a document changes the model's internal representation of a query: a relevant document should shift the model's internal representation characteristically. This observation leads us to formalize and study \textbf{representational shifts}, the change in the model's representation of the query induced by a document in context. In Sec.~\ref{sec:methodology}, we show that the geometry of RS encodes relevance information that, when transformed through a learned projection can output scores that are calibrated towards a natural decision boundary.

\subsection{Contributions}
\label{sec:contributions}

Our key contributions are as follows: We highlight the \textbf{threshold inconsistency} problem in current SOTA rerankers and provide a means to quantify its impact. We identify \textbf{representational shifts} as a relevance signal: changes in the query's value vectors induced by conditioning on a document in context. We introduce a \textbf{lightweight learning framework} that maps the representational shift space to calibrated scores, yielding a consistent decision boundary across datasets. We demonstrate \textbf{competitive performance} across six diverse retrieval datasets, achieving 2.0 and 7.2-point gains in Recall@5 and F1 at the natural threshold relative to baselines, while using only 2.3M trained parameters on top of frozen LLM representations. 

\section{Threshold Inconsistency in Rerankers}
\label{sec:motivation}

We evaluate Qwen-Reranker-8B~\citep{qwen3embedding} on six retrieval datasets (Sec.~\ref{sec:setup}) and analyze the resulting score distributions to show the threshold inconsistency problem.

\subsection{Optimal Thresholds Vary Across Datasets}

\begin{figure}[t]
\centering
\includegraphics[width=\linewidth]{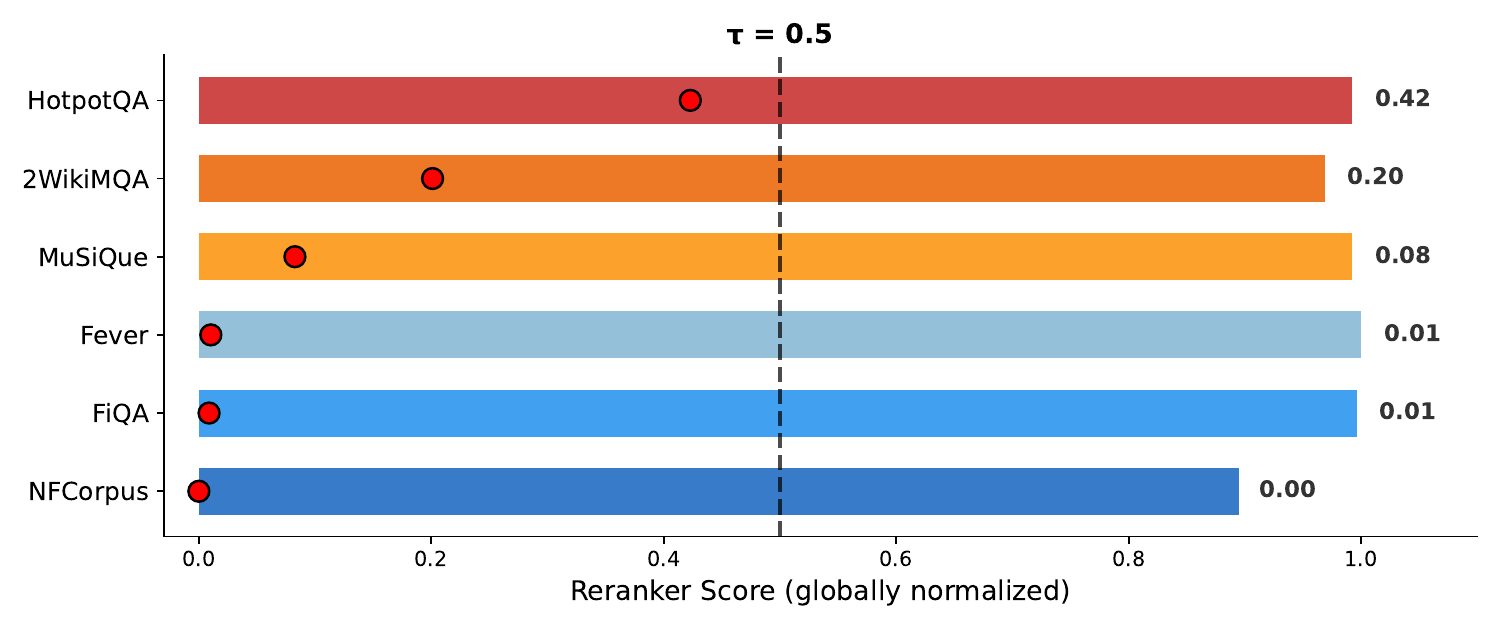}
\caption{Optimal threshold for Qwen3-Reranker-8B for F1 across datasets. The $x$-axis shows the range of scores (globally normalized); the optimal threshold is indicated by the red dot.}
\label{fig:score_range_qwen}
\end{figure}

For each dataset, we compute the \emph{optimal threshold}, the threshold achieving the highest mean F1, across 500 queries. We plot this against the per-dataset score range after applying a global min-max normalization (mapping the score range to $[0,1]$) so that models with different native scales can be compared directly. To quantify dataset-level calibration we report two metrics: \textbf{Bias}, the absolute offset between the optimal threshold and the model's natural decision boundary ($\tau$), and \textbf{Variance}, the spread of per-dataset optimal thresholds around their mean. Fig.~\ref{fig:score_range_qwen} shows the results for Qwen3-Reranker-8B: bias=0.379, variance=0.023. This reveals that the natural threshold ($\tau{=}0.5$) consistently overshoots the true decision boundary, while the optimal threshold varies substantially across domains. Consequently, effective deployment requires labeled data for calibration, limiting out-of-the-box performance.

\subsection{Fixed Thresholds Hurt Individual Queries}

\begin{figure}[t]
\centering
\includegraphics[width=\columnwidth]{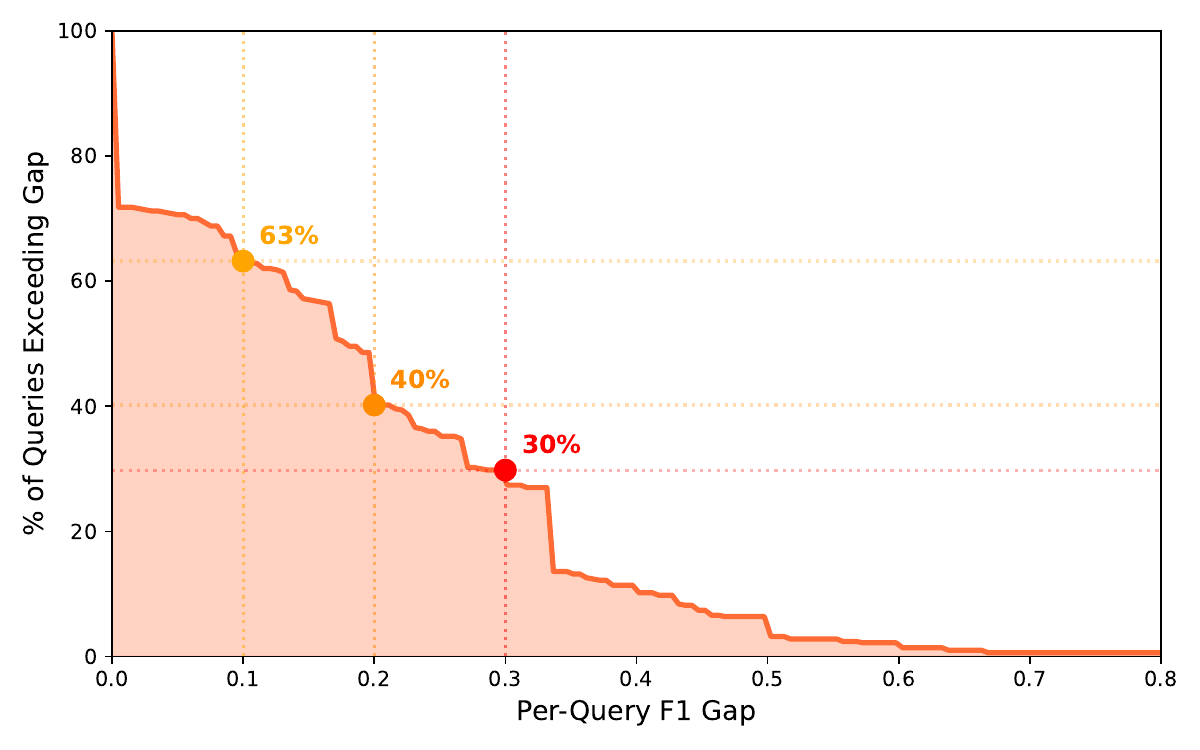}
\caption{F1 gap CDF for Qwen3-Reranker-8B on HotpotQA. The $x$-axis shows the F1 gap between the dataset-level optimal threshold and the per-query optimal threshold; the $y$-axis shows the fraction of queries exceeding that gap. 63\% of queries lose $\geq$0.1 F1 from using a fixed threshold.}
\label{fig:f1_gap}
\end{figure}

Even within a single dataset, the optimal threshold varies substantially across queries. Fig.~\ref{fig:f1_gap} illustrates this effect on HotpotQA by showing the fraction of queries for which the F1 score obtained using the dataset-level optimal threshold falls short of the per-query optimal F1. For Qwen3-Reranker-8B, 63\% of queries incur an F1 loss greater than 0.1, and 30\% incur a loss greater than 0.3. 

\begin{table}[t]
\centering
\caption{Paired $t$-test: per-query optimal F1 vs.\ dataset-optimal threshold F1 for Qwen3-Reranker-8B.}
\label{tab:ttest_qwen8b}
\footnotesize
\setlength{\tabcolsep}{3pt}
\begin{tabular}{lrcccrl}
\toprule
\textbf{Dataset} & \textbf{$N$} & \textbf{Q-Opt} & \textbf{D-Opt} & \textbf{Gap} & \textbf{$t$} & \textbf{$p$-value} \\
\midrule
2WikiMQA  & 500 & 73.0 & 55.0 & 18.1 & 24.22 & $<10^{-99}$ \\
Fever     & 500 & 100.0 & 99.5 &  0.5 &  2.73 & $6.5{\times}10^{-3}$ \\
FiQA      & 500 &  99.1 & 95.0 &  4.0 &  8.05 & $6.0{\times}10^{-15}$ \\
HotpotQA  & 500 &  79.7 & 61.0 & 18.8 & 23.94 & $<10^{-99}$ \\
MuSiQue   & 500 &  77.7 & 61.0 & 16.7 & 22.93 & $<10^{-99}$ \\
NFCorpus  & 323 &  82.1 & 64.9 & 17.2 & 15.40 & $<10^{-99}$ \\
\bottomrule
\end{tabular}
\end{table}

To quantify performance loss attributable specifically to poor calibration rather than reranking quality, we conduct a paired t-test comparing dataset-level optimal F1 scores with per-query optimal F1 scores. Table~\ref{tab:ttest_qwen8b} reports the results for Qwen3-Reranker-8B across datasets. In all cases, the difference between the two scores is statistically significant ($p < 0.05$), indicating that a large portion of the observed performance gap arises from calibration error rather than limitations in ranking ability.

\section{Related Work}
\label{sec:related}

\paragraph{Ranking paradigms.} Reranking methods can be broadly categorized into three paradigms. \emph{Pointwise} methods score each query--document pair independently, enabling efficient threshold-based selection~\cite{nogueira-etal-2020-document, DBLP:conf/sigir/MaWYWL24, qwen3embedding}. \emph{Listwise} methods condition on the entire candidate set and directly optimize or generate ranked permutations~\cite{pradeep2023rankzephyr, sun-etal-2023-chatgpt}. \emph{Setwise} methods iteratively identify the most relevant document from subsets of candidates~\cite{10.1145/3696410.3714863,10.1145/3626772.3657813}. Since listwise and setwise approaches incur substantially higher computational and latency overheads for large candidate sets, we primarily compare against pointwise rerankers.

\paragraph{Model architectures.} Reranking models fall into three architectural families. \emph{Cross-encoders} jointly encode the query and document for relevance prediction~\cite{nogueira-etal-2020-document, DBLP:journals/corr/abs-2101-05667, colbert2020khattab}. \emph{Open-source LLMs} have been adapted for pointwise, listwise, and setwise reranking~\cite{DBLP:conf/sigir/MaWYWL24, pradeep2023rankzephyr, qwen3embedding, DBLP:journals/corr/abs-2508-09497, sun2026rethinkingrerankerboundaryawareevidence, behnamghader2026llm2vecgengenerativeembeddingslarge}. \emph{Closed-source LLMs} are often used as zero-shot rerankers via prompting~\cite{sun-etal-2023-chatgpt}. Across these families, relevance is typically inferred from language modeling objectives rather than explicit relevance supervision. We primarily compare against cross-encoder and open-source LLM rerankers of equivalent scale.

\paragraph{Threshold Calibration.} Prior work studies calibration techniques for making reranker scores comparable across queries and suitable for threshold-based decisions. Methods based on Platt scaling and related calibration approaches~\cite{platt1999probabilistic, posokhov-etal-2025-relevance, 10.1145/3696410.3714658, yu-etal-2025-explain} convert raw scores into calibrated confidence estimates, but still require task-specific calibration and externally defined thresholds. Other approaches derive statistically grounded thresholds~\cite{li-etal-2022-certified} or use predictive uncertainty for selective acceptance~\cite{11159229, yoon2025acurank}, yet they also depend on auxiliary decision rules. RSRank learns a consistent relevance boundary directly during training, while remaining compatible with post-hoc calibration and thresholding techniques.

\paragraph{Probing LLM Representations.} Recent work shows that intermediate LLM representations encode task-relevant signals. Intermediate layers improve embedding quality~\citep{skean2025layer}, contrastive layer analysis enhances factuality~\citep{zhang2024sled}, hidden states support document attribution~\citep{phukan-etal-2024-peering}, and attention patterns have been used for reranking~\citep{chen2025attention}. These findings highlight the importance of leveraging model internals beyond next-token prediction. In contrast to prior work on hidden states or attention weights, our approach uses value vector shifts to capture how a document updates the model’s internal representation of a query, directly aligning the representation with the relevance decision.

\paragraph{Intrinsic Geometry in LLMs.} Prior work shows that neural representations are highly anisotropic, often forming a cone-shaped geometry in embedding space~\citep{ait-saada2023anisotropy}, a property observed consistently across layers and architectures~\citep{razzhigaev2024shape, skean2025layer}. Rather than being a training artifact, this geometry has been argued to encode meaningful semantic and structural information~\citep{godey2024anisotropy, kudrjashov2025shrink}. Although some methods attempt to suppress anisotropy through normalization or whitening, recent evidence suggests that doing so can degrade generation and downstream performance, implying that anisotropy itself carries useful signals~\citep{godey2024anisotropy, kudrjashov2025shrink}. These observations motivate our approach: we leverage the anisotropic structure of RS to extract a relevance-discriminative signal.

\section{Methodology}
\label{sec:methodology}

\subsection{Finite Difference as Representational Shift}
\label{sec:finite-diff-features}

We consider a decoder-only Transformer with $L$ layers and $H$ attention heads per layer.
Let $\Sigma$ denote the vocabulary and let a document and query be token sequences
$d=(d_1,\ldots,d_n)\in\Sigma^n$ and $q=(q_1,\ldots,q_m)\in\Sigma^m$, respectively.
We study how the document prefix alters internal representations of query tokens during \emph{prefill}.

\paragraph{Pre-attention value vectors.}
Fix a layer $\ell\in[L]$ and head $h\in[H]$ with head dimension $d_h$.
Let $r^{(\ell-1)}_i(x)\in\mathbb{R}^{D}$ denote the residual stream vector entering layer $\ell$ at
position $i$ for input sequence $x$, and let $W^{(\ell,h)}_V\in\mathbb{R}^{d_h\times D}$ be the
value projection matrix for head $(\ell,h)$. The pre-attention value vector at position $i$ is
\begin{equation}
v^{(\ell,h)}_i(x) := W^{(\ell,h)}_V \, r^{(\ell-1)}_i(x)\in\mathbb{R}^{d_h}.
\end{equation}
In our reranking setup, we focus on value vectors of the query tokens in the
concatenated input $x=d\cdot q$.

\paragraph{Controlling for prefix length.}
Prepending a document changes both (i) the \emph{content} available for attention and
(ii) the \emph{absolute positions} of query tokens.
Since our goal is to isolate only \emph{prefix content effects}, we need to compare the document prefix to a
length-matched \emph{null prefix} $\varnothing_n\in\Sigma^n$ that carries minimal semantic content
(e.g., padding or benign filler tokens), yielding a controlled finite-difference feature.
Specifically, for each query token position $t\in[m]$, we define the document-induced delta of the value vector at head $(\ell,h)$ as
\begin{equation}
\delta v^{(\ell,h)}_t(d;q)
:= v^{(\ell,h)}_{n+t}(d\cdot q) - v^{(\ell,h)}_{n+t}(\varnothing_n\cdot q)
\in\mathbb{R}^{d_h}
\label{eq:delta-v}
\end{equation}

a standard construction in discrete/finite-difference calculus (Appendix~\ref{app:discrete-calculus}). However, specifically for value vector based signals, we can simplify this construction to Eq.~(\ref{eq:delta-v-rope}) when our base model uses RoPE~\citep{10.1016/j.neucom.2023.127063}, which encodes relative positions. Under RoPE, the QK attention between tokens of the query remains the same regardless of how far the query is shifted, and since the value vectors themselves are not subject to RoPE, they are not affected by their position.

\begin{equation}
\delta v^{(\ell,h)}_t(d;q)
:= v^{(\ell,h)}_{n+t}(d\cdot q) - v^{(\ell,h)}_{n+t}(q)
\in\mathbb{R}^{d_h}
\label{eq:delta-v-rope}
\end{equation}

\paragraph{Representational shift tensor.}
Let $\mathcal{I}\subseteq[L]\times[H]$ be a selected set of $L$ layers and $H$ heads.
Collecting the per-head deltas across query token positions $t\in[m]$, we define the representational shift tensor $\boldsymbol{\Delta}(d,q) \in \mathbb{R}^{L \times H \times T \times D}$ as
\begin{equation}
\boldsymbol{\Delta}(d,q)_{\ell,h,t} := \delta v^{(\ell,h)}_t(d;q)
\label{eq:delta-tensor}
\end{equation}
where $D \equiv d_h$ is the head dimension.
Intuitively, $\boldsymbol{\Delta}(d,q)$ captures how the document prefix \emph{re-contextualizes} each query token in value space, at a resolution indexed by $(\ell,h,t)$.
Given a set of documents $\mathcal{D}$, we write $\boldsymbol{\Delta}(\mathcal{D},q)$ for the shift induced by conditioning on all documents in $\mathcal{D}$.

\subsection{Representational Shift Models Relevance}
\label{sec:rs-motivation}

\paragraph{Notation.}
For a query $q$ with candidate document set $\mathcal{D}_q$, let $\mathcal{D}^+_q \subseteq \mathcal{D}_q$ denote the set of relevant and $\mathcal{D}^-_q = \mathcal{D}_q \setminus \mathcal{D}^+_q$ the irrelevant documents.

\paragraph{Oracle shift.}
We define the \emph{oracle shift} $\boldsymbol{\Delta}^*(q) := \boldsymbol{\Delta}(\mathcal{D}^+_q, q)$ as the representational shift induced by conditioning on \emph{all} relevant documents simultaneously.
This shift encodes the aggregate effect that the complete relevant context has on the model's internal representations of the query.

\paragraph{Oracle-similarity ranking.}
For each candidate document $d_i \in \mathcal{D}_q$, we compute the cosine similarity between its individual shift $\boldsymbol{\Delta}(d_i, q)$ and the oracle shift $\boldsymbol{\Delta}^*(q)$, and rank documents accordingly.
On the 2WikiMQA validation set, this oracle-similarity ranking achieves R@5\,=\,89.5, P@5\,=\,42.8, and F1@5\,=\,57.9---demonstrating that alignment with oracle is effective in separating relevant and irrelevant documents, and that the geometry of the shift space encodes relevance structure.

\paragraph{From oracle similarity to learned projection.}
The oracle shift is unavailable at inference time.
However, the experiment above suggests that if we can learn a projection $\mathbf{B}$ that transforms the RS space such that $\mathbf{B}\,\boldsymbol{\Delta}^*(q)$ falls in a chosen orthant specifically aligning to $\mathbf{1}_P$ across queries, then the oracle similarity $\cos\!\big(\mathbf{B}\,\boldsymbol{\Delta}(d, q),\; \mathbf{B}\,\boldsymbol{\Delta}^*(q)\big)$ reduces to $\cos\!\big(\mathbf{B}\,\boldsymbol{\Delta}(d, q),\; \mathbf{1}_P\big)$, applicable at inference.
This motivates our learning framework.

\subsection{Learning Calibrated Projections}
\label{sec:learning}

\subsubsection{Projection Matrix}

We learn a projection matrix $\mathbf{B} \in \mathbb{R}^{L \times H \times P \times D}$, where $P$ is the projection dimension. For each layer $\ell$, head $h$, the submatrix $\mathbf{B}_{\ell,h} \in \mathbb{R}^{P \times D}$ projects the $D$-dimensional shift into a $P$-dimensional space.

\paragraph{Scoring Function.} Given a candidate document $d$, we compute its relevance score as:
\begin{align}
\mathbf{z}_{\ell,h,t} &= \mathbf{B}_{\ell,h} \, \boldsymbol{\Delta}(d, q)_{\ell,h,t} \in \mathbb{R}^P \\
s(d \mid q) &= \sum_{\ell,h} \frac{1}{T} \sum_{t=1}^{T} \cos(\mathbf{z}_{\ell,h,t}, \mathbf{1}_P)
\end{align}
where $\mathbf{1}_P$ denotes the all-ones vector in $\mathbb{R}^P$. Intuitively, $\mathbf{B}$ learns to extract a ``relevance direction'' from representational shift vectors: documents whose projected shifts have high cosine similarity with $\mathbf{1}_P$ receive high scores.

\subsubsection{Training Objectives}

Our training objective consists of five terms designed to achieve calibrated separation of relevant and irrelevant documents at a fixed threshold.

\paragraph{Calibration Loss.} The core objective pushes relevant documents to have positive scores and irrelevant documents to have negative scores:
\begin{equation}
\mathcal{L}_{\text{cal}} = \frac{1}{|\mathcal{D}^+|} \sum_{d \in \mathcal{D}^+} [\text{-}s(d)]_+ + \frac{1}{|\mathcal{D}^-|} \sum_{d \in \mathcal{D}^-} [s(d)]_+
\end{equation}
where $[\cdot]_+ = \max(\cdot, 0)$ denotes the ReLU function and $\mathcal{D}^+$, $\mathcal{D}^-$ are the relevant and irrelevant document sets as defined in Sec.~\ref{sec:rs-motivation}. This loss creates a natural decision boundary at $s = 0$.

\paragraph{Margin Loss.} To ensure robust separation, we enforce a margin $m$ between classes:
\begin{align}
\mathcal{L}_{\text{margin}} &= \frac{1}{|\mathcal{D}^+|} \sum_{d \in \mathcal{D}^+} [m - s(d)]_+ \nonumber \\
&\quad + \frac{1}{|\mathcal{D}^-|} \sum_{d \in \mathcal{D}^-} [s(d) + m]_+ \nonumber \\
&\quad + \Big[m - \big(\min_{d \in \mathcal{D}^+} s(d) - \max_{d \in \mathcal{D}^-} s(d)\big)\Big]_+
\end{align}
The first two terms push relevant scores above $+m$ and irrelevant scores below $-m$. The third term ensures that the lowest-scoring relevant document exceeds the highest-scoring irrelevant document by at least $m$. We use $m = 0.5$ in all experiments.

\paragraph{Orthogonality Regularization.} To prevent dimension collapse in the projection, we regularize each $\mathbf{B}_{\ell,h}$ to have orthonormal rows:
\begin{equation}
\mathcal{L}_{\text{ortho}} = \frac{1}{LH} \sum_{\ell,h} \frac{\|\mathbf{B}_{\ell,h} \mathbf{B}_{\ell,h}^\top - \mathbf{I}_P\|_F^2}{P^2}
\end{equation}
This ensures all $P$ dimensions are effectively utilized and prevents the projection from degenerating to a lower-rank mapping.

\paragraph{Oracle Alignment.} We provide explicit supervision on the direction of the oracle shift projection:
\begin{equation}
\mathcal{L}_{\text{align}} = 1 - \mathbb{E}_{\ell,h,t}\big[\cos(\mathbf{B}_{\ell,h} \, \boldsymbol{\Delta}^*(q)_{\ell,h,t}, \mathbf{1}_P)\big]
\end{equation}
where $\boldsymbol{\Delta}^*(q) = \boldsymbol{\Delta}(\mathcal{D}^+, q)$ is the oracle representational shift induced by the full set of relevant documents (Sec.~\ref{sec:rs-motivation}). This anchors the ``ideal relevance direction'' to the ones vector.

\paragraph{Magnitude Constraint.} For training stability, we bound the Frobenius norm of $\mathbf{B}$:
\begin{equation}
\mathcal{L}_{\text{frob}} = \big[\|\mathbf{B}\|_F - c\big]_+, \quad c = \tfrac{1}{2}\sqrt{LHP D}
\end{equation}

\paragraph{Total Objective.} The complete training loss is:
\begin{equation}
\mathcal{L} = \lambda_1 \mathcal{L}_{\text{cal}} + \mathcal{L}_{\text{margin}} + \lambda_2 \mathcal{L}_{\text{ortho}} + \lambda_3 \mathcal{L}_{\text{align}} + \lambda_4 \mathcal{L}_{\text{frob}}
\end{equation}

\section{Experimental Setup}
\label{sec:setup}

\subsection{Baseline}
\label{sec:baseline}
We compare against Qwen3-Reranker-8B~\citep{qwen3embedding}, a state-of-the-art LLM reranker built on Qwen3-8B~\citep{qwen3technicalreport} that produces calibrated binary relevance scores. For fairness, RSRank uses frozen representations from the same backbone. We focus on Qwen3-Reranker-8B because it substantially outperforms earlier rerankers such as BGE-Reranker~\citep{chen-etal-2024-m3} and Jina Reranker v2~\citep{jinaaiji29:online}. We additionally compare against MonoT5~\citep{nogueira-etal-2020-document}, a standard cross-encoder baseline, and LLM2Vec-Gen~\citep{behnamghader2026llm2vecgengenerativeembeddingslarge} (Qwen3-8B) to isolate gains beyond representation quality alone.

\subsection{Datasets}
\label{sec:datasets}

We evaluate on six retrieval datasets spanning multi-hop reasoning and domain specific retrieval.

\paragraph{Multi-hop QA.}
\emph{2WikiMultihopQA}~\citep{ho-etal-2020-constructing} focuses on compositional reasoning across pairs of Wikipedia articles with sentence-level supporting facts. \emph{HotpotQA}~\citep{yang-etal-2018-hotpotqa} emphasizes multi-hop reasoning in a distractor setting, where each question is paired with 2 supporting and 8 TF-IDF-retrieved paragraphs. \emph{MuSiQue}~\citep{trivedi-etal-2022-musique} increases reasoning complexity by composing multiple single-hop questions into multi-hop questions and includes adversarial unanswerable examples.

\paragraph{Domain-Specific Retrieval.}
\emph{FiQA}~\citep{fiqa2018maia} contains opinion-based financial question answering data. \emph{FEVER}~\citep{thorne-etal-2018-fever} is a fact verification dataset where claims must be supported or refuted using evidence from Wikipedia. \emph{NFCorpus}~\citep{boteva2016nfcorpus} is a biomedical retrieval dataset linking natural-language nutrition and medical queries to scientific documents.

\paragraph{Document granularity.}
The multi-hop QA datasets (2WikiMQA, HotpotQA) operate at sentence-level granularity, while BEIR datasets and MuSiQue use paragraph-level documents. In practice, RAG pipelines operate on chunked passages granularities, which is the primary setting we target. Our method is trained and evaluated jointly on these mixed chunk lengths, demonstrating generalisation across the passage sizes typical of chunked retrieval. Reranking over longer-form documents (e.g., entire articles) is an interesting direction but falls outside the scope of this work.

\subsection{Training}
\label{sec:training}

\paragraph{Training protocol.}
RS features are pre-computed by running a forward pass per query-document pair.
We train the projection matrix $\mathbf{B}$ (2.3M parameters; Sec.~\ref{sec:learning}) on just \textbf{2000} samples from the datasets listed above (excluding Fever, which serves as a zero-shot test).
Layer~0 is excluded from training since its value vectors are raw token embeddings rather than contextualized representations. Complete training hyperparameters are provided in Appendix~\ref{app:training-details}.

\paragraph{Training cost.}
RSRank has a distinct advantage in the training phase compared to the regular finetuning carried out in Qwen3-Reranker-8B because it works on top of frozen LLM representations, training an independent projection matrix. Because of this, the expensive forward pass of the LLM has to be done only once and the backpropagation does not need to update the parameters of the LLM.

\subsection{Evaluation}
\label{sec:evaluation}

\paragraph{Evaluation protocol.}
For each dataset, we evaluate on up to 500 sampled queries from the test split.
For multi-hop datasets, each query comes with its original set of gold and distractor documents.
For BEIR datasets, we pair each query with its relevant documents plus 15 randomly sampled negatives from the corpus.

\paragraph{Metrics.} We report the following metrics:
\begin{enumerate}[leftmargin=*,itemsep=0pt,topsep=1pt,parsep=0pt,partopsep=0pt]
    \item \textbf{NDCG@5}: Normalized Discounted Cumulative Gain at rank 5, measuring ranking quality.
    \item \textbf{Recall@5}: Fraction of relevant documents appearing in the top 5 ranked positions, measuring retrieval coverage.
    \item \textbf{F1@$\boldsymbol{\tau}$}: F1 score computed by thresholding scores at each method's \emph{natural} decision boundary. This metric measures how well a method separates relevant from irrelevant documents \emph{without dataset-specific tuning}, and is our primary metric for evaluating calibration quality.
\end{enumerate}

\paragraph{Inference cost.}
RSRank requires an additional query-only forward pass over baselines to compute the null-prefix baseline, followed by a lightweight projection and cosine similarity step. However, because the query-only pass is amortized across all documents for a query, the effective per-document overhead is negligible. As shown in Table~\ref{tab:prefill_overhead}, the full RSRank pipeline adds only 15.4\,ms (+0.88\%) over standard query+document forwards on an A100-80GB GPU.

\begin{table}[t]
\centering
\caption{Time to rerank 100 docs on A100-80GB GPU (mean$\pm$std over 30 runs) for Qwen3-8B. \emph{100 fwd}: query+document prompt batched. \emph{101 fwd}: + one query-only prompt. \emph{101 fwd + score}: full RSRank pipeline.
}
\label{tab:prefill_overhead}
\small
\setlength{\tabcolsep}{3pt}

\resizebox{\columnwidth}{!}{%
\begin{tabular}{lccc}
\toprule
\textbf{Stage} & \textbf{Prompt length} & \textbf{Time (ms)} & \textbf{Overhead} \\
\midrule
100 fwd & $100 \times 229$ & $1747.2 \pm 1.3$ & --- \\
101 fwd         & $+ 1 \times 82$  & $1749.9 \pm 1.2$ & $+2.7$ ($+0.15\%$) \\
101 fwd + scoring & same                  & $1762.6 \pm 6.4$ & $+15.4$ ($+0.88\%$) \\
\bottomrule
\end{tabular}%
}

\end{table}

\section{Results}
\label{sec:results}

\begin{table*}[t]
\caption{Results across six retrieval datasets. Best result per column in \textbf{bold}; ties within 1 point are co-bolded. F1@$\tau$ uses each method's default decision boundary without dataset-specific tuning.}
\label{tab:main_results}
\centering
\small
\setlength{\tabcolsep}{4pt}
\begin{tabular}{l ccccccc}
\toprule
    \textbf{Method} & \textbf{2WikiMQA} & \textbf{HotpotQA} & \textbf{MuSiQue} & \textbf{FiQA} & \textbf{Fever} & \textbf{NFCorpus} & \textbf{Avg} \\
\midrule
\multicolumn{8}{l}{\emph{NDCG@5 $\uparrow$}} \\
MonoT5-base (0.2B)      & 61.8 & 68.6 & 66.7 & 96.5 & 99.9 & 80.0 & 78.9 \\
MonoT5-3B               & 64.4 & 73.7 & 73.5 & 98.5 & \textbf{100.0} & 84.9 & 82.5 \\
LLM2Vec-Gen (Qwen3-8B)  & 55.7 & 60.0 & 62.9 & 97.0 & 99.7 & 83.6 & 76.5 \\
Qwen3-Reranker-8B       & 71.6 & \textbf{81.1} & 78.8 & \textbf{99.4} & \textbf{100.0} & 86.9 & 86.3 \\
RSRank (Ours)           & \textbf{80.2} & 79.6 & \textbf{84.0} & 97.5 & 99.2 & 83.4 & \textbf{87.3} \\
\midrule
\multicolumn{8}{l}{\emph{Recall@5 $\uparrow$}} \\
MonoT5-base (0.2B)      & 63.0 & 71.6 & 68.6 & 94.4 & 99.8 & 31.0 & 71.4 \\
MonoT5-3B               & 66.0 & 76.4 & 76.7 & 96.3 & 99.9 & 34.6 & 75.0 \\
LLM2Vec-Gen (Qwen3-8B)  & 60.5 & 63.9 & 69.8 & 95.1 & \textbf{99.8} & 35.1 & 70.7 \\
Qwen3-Reranker-8B       & 74.0 & \textbf{84.3} & 84.0 & \textbf{96.7} & \textbf{99.8} & 35.8 & 79.1 \\
RSRank (Ours)           & \textbf{85.1} & 83.0 & \textbf{88.8} & \textbf{95.6} & 99.7 & 34.3 & \textbf{81.1} \\
\midrule
\multicolumn{8}{l}{\emph{F1@$\tau$ (natural threshold) $\uparrow$}} \\
MonoT5-base \scriptsize{($\tau{=}0.5$)}            & 45.9 & 47.3 & 49.0 & 53.9 & 91.3 & 11.5 & 49.8 \\
MonoT5-3B \scriptsize{($\tau{=}0.5$)}              & 48.1 & 53.3 & 55.4 & 62.2 & 91.5 &  9.1 & 53.3 \\
LLM2Vec-Gen (Qwen3-8B) \scriptsize{($\tau{=}0$)}              & 16.7 & 11.9 & 18.2 & 23.8 & 13.1 & \textbf{61.5} & 24.2 \\
Qwen3-Reranker-8B \scriptsize{($\tau{=}0.5$)}      & 51.9 & \textbf{60.6} & 57.1 & 80.6 & \textbf{98.2} & 13.5 & 60.3 \\
RSRank (Ours) \scriptsize{($\tau{=}0$)}            & \textbf{60.5} & 51.8 & \textbf{61.2} & \textbf{85.9} & 83.4 & \textbf{62.2} & \textbf{67.5} \\
\bottomrule
\end{tabular}
\end{table*}

\subsection{Ranking Quality}
\label{sec:ranking_quality}
As shown in Table~\ref{tab:main_results}, RSRank achieves the best average NDCG@5 (87.3) and Recall@5 (81.1) across all six datasets. The largest gains appear on the multi-hop benchmarks 2WikiMQA and MuSiQue, where RSRank outperforms Qwen3-Reranker-8B by 8.6 and 5.2pp in NDCG@5, respectively. On HotpotQA and the BEIR benchmarks, performance is largely comparable, with Qwen3-Reranker-8B holding a slight edge on HotpotQA and NFCorpus. Both methods achieve near-perfect results on FEVER and FiQA, while NFCorpus remains challenging for both. These results demonstrate that representational shifts provide a competitive alternative to fully trained rerankers.

\subsection{Selection Quality}
\label{sec:selection_quality}

RSRank, evaluated at its designed threshold of $\tau{=}0$, achieves the highest average F1 score (67.5), outperforming Qwen3-Reranker-8B at its default threshold of $\tau{=}0.5$ (60.3) by 7.2pp. The largest gap appears on NFCorpus, where Qwen3-Reranker-8B is severely miscalibrated: its default threshold yields an F1 of only 13.5, despite the dataset-optimal threshold lying near zero. RSRank also shows consistent gains on 2WikiMQA, MuSiQue, and FiQA, suggesting that representational shifts provide a more robust relevance signal under fixed-threshold evaluation.

\begin{figure}[t]
\centering
\includegraphics[width=\linewidth]{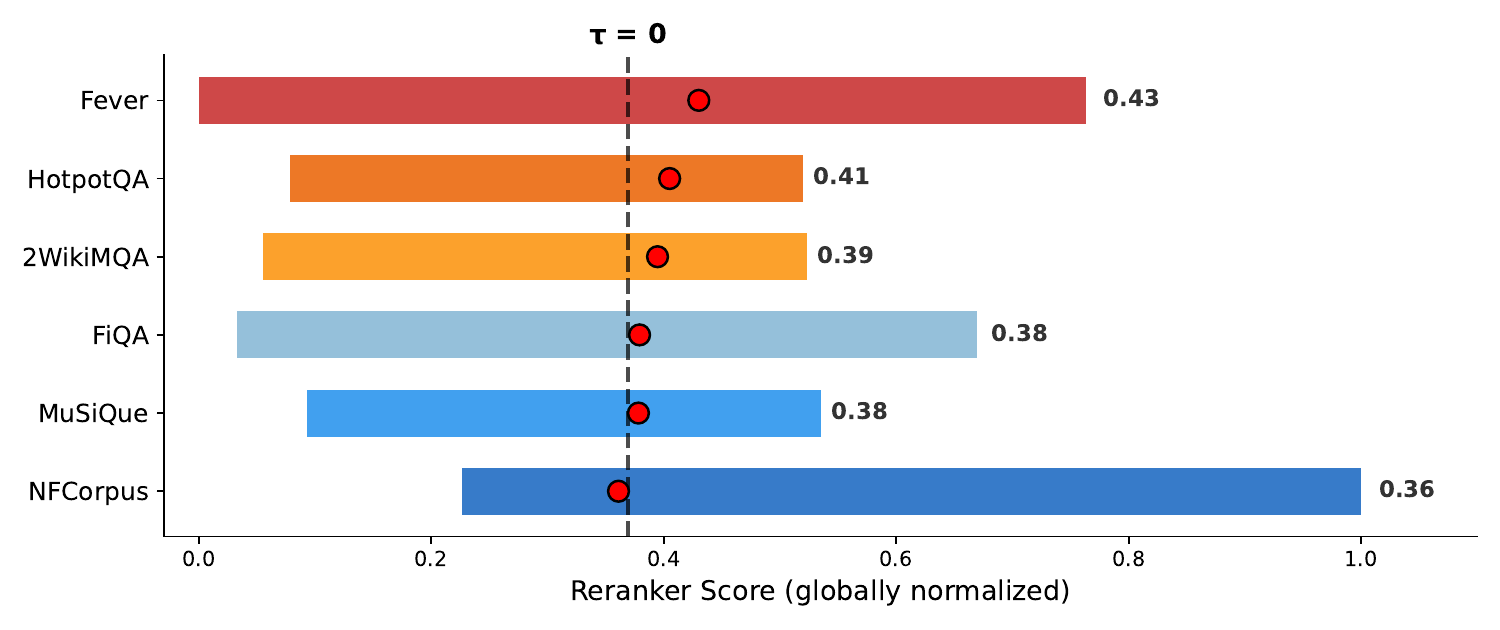}
\caption{Optimal threshold for RSRank for best mean F1 across datasets. The $x$-axis shows the range of scores (globally normalized) given by the reranker; the optimal threshold is indicated by the red dot. Optimal threshold Bias: 0.0221, Variance: 0.0005.}
\label{fig:score_range_our}
\end{figure}

\subsection{Threshold Stabilization Across Datasets}
\label{sec:dataset-thr-stable}
RSRank produces substantially more consistent thresholds across datasets than Qwen3-Reranker-8B, reducing threshold bias from 0.379 to 0.022 (17$\times$ lower) and variance from 0.023 to 0.0005 (47$\times$ lower). Fig.~\ref{fig:score_range_our} shows that RS training aligns per-dataset optimal thresholds under a shared global normalization, improving calibration robustness across datasets.

\section{Analysis}
\label{sec:analysis}

We conduct ablation studies on the 2WikiMQA validation set to understand which design choices drive RSRank's performance. Ablations on architectural choices, loss components and comparisons with analytical baselines are presented in Appendix.~\ref{app:Ablations}  

\subsection{Separability of Representational Shift}
\label{sec:separability}

\begin{figure}[t]
  \centering
  \includegraphics[width=\linewidth]{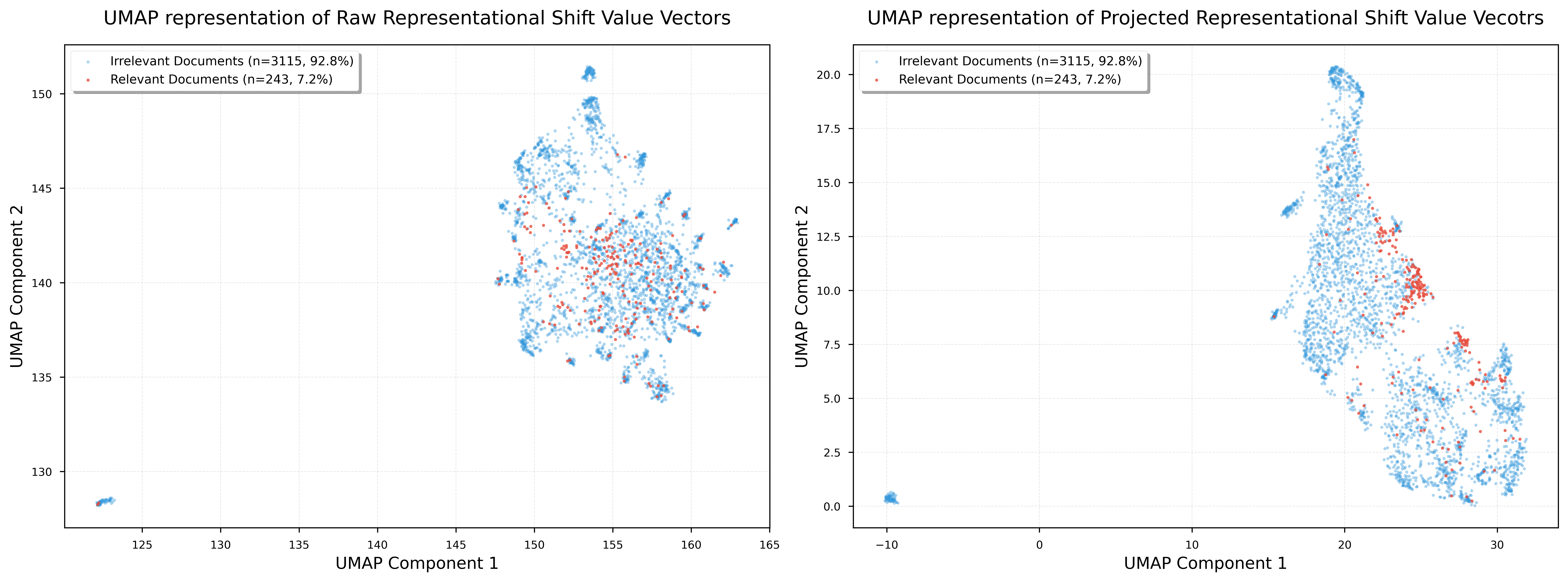}
  \caption{
    UMAP visualization of representational shifts on 2WikiMQA.
    \textbf{Left:} Raw shifts, where relevant (red) and irrelevant (blue) documents overlap.
    \textbf{Right:} After projection with $\mathbf{B}$, relevant documents form a clearly separated cluster.
  }
  \label{fig:umap}
\end{figure}

Fig.~\ref{fig:umap} visualizes RS vectors via UMAP for 100 queries (3115 irrelevant, 243 relevant documents). Raw shifts (left) show relevant and irrelevant documents thoroughly intermixed---the shift signal alone does not linearly separate classes. After the learned projection $\mathbf{B}$ (right), relevant documents consolidate into a distinct cluster, confirming that $\mathbf{B}$ extracts a relevance-discriminative subspace.

\subsection{Sample Efficiency}
\label{sec:sample_efficiency}

The stability of the RS space allows for the learned projection to converge with remarkably few examples. Table~\ref{tab:sample_efficiency} shows that just 50 samples achieve 89.2\% R@5 ($<$5pp of the 800-sample model), and performance saturates around 400 samples.

\begin{table}[t]
\centering
\caption{Sample efficiency on 2WikiMQA (10 epochs, Qwen3-8B). Performance saturates by 400 samples.}
\label{tab:sample_efficiency}
\small
\begin{tabular}{lcccccc}
\toprule
\textbf{Samples} & 0 & 50 & 100 & 200 & 400 & 800 \\
\midrule
\textbf{R@5}  & 28.7 & 89.2 & 90.7 & 91.1 & 93.8 & 93.9 \\
\textbf{F1@0} & 13.2 & 74.5 & 76.9 & 77.7 & 80.8 & 82.1 \\
\bottomrule
\end{tabular}
\end{table}

\subsection{Per-Query Calibration}
\label{sec:per-query-gap}

As established in Sec.~\ref{sec:motivation}, threshold calibration operates at two levels: \emph{dataset-level} and \emph{query-level}. Sec.~\ref{sec:dataset-thr-stable} showed that RSRank effectively addresses dataset-level calibration. We now examine the query-level picture.

\begin{figure}[t]
\centering
\includegraphics[width=\columnwidth]{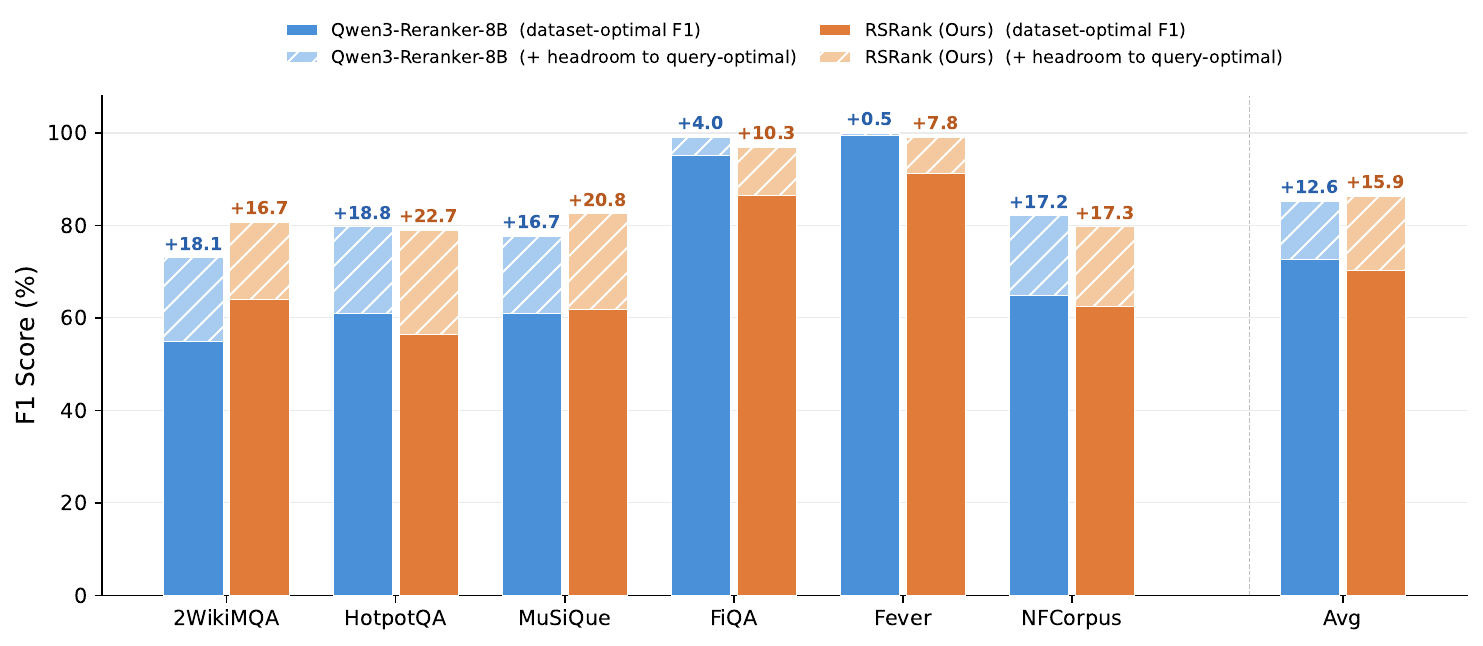}
\caption{Dataset-optimal F1 and headroom to per-query optimal F1 for Qwen3-Reranker-8B and RSRank. RSRank achieves higher per-query optimal F1 on average (86.3 vs.\ 85.3), indicating better ranking quality}
\label{fig:scope_for_improvement}
\end{figure}

Fig.~\ref{fig:scope_for_improvement} decomposes each model's F1 into two components: the dataset-optimal F1 and the additional headroom to per-query optimal F1. RSRank achieves a higher per-query optimal F1 than Qwen3-Reranker-8B on average, indicating stronger underlying ranking quality. The headroom from dataset-optimal to query-optimal is larger for RSRank (+15.9 vs.\ +12.6 on average). These results indicate that RSRank provides a strong foundation with better dataset-level calibration. The superior per-query ranking of RSRank indicates that future work on per-query calibration can further improve the performance.

\section{Conclusion}
\label{sec:conclusion}

We present RSRank, a reranking method that uses representational shifts (RS) of value vectors to produce relevance scores calibrated at a natural decision boundary. Our motivation identifies two levels of threshold calibration---dataset-level and query-level---and shows how existing rerankers suffer in both areas. RSRank addresses dataset-level calibration by learning a lightweight projection (2.3M params) that maps RS to scores, reducing dataset-level threshold bias by $17{\times}$ and variance by $47{\times}$ compared to baselines. Across six diverse retrieval benchmarks, RSRank achieves the highest average NDCG@5, Recall@5, and F1 at its natural threshold, outperforming SOTA Qwen3-Reranker-8B. A headroom analysis further shows that RSRank attains higher per-query optimal F1 than the baseline, confirming stronger underlying ranking quality.

\paragraph{Future Work.}
The remaining headroom from dataset to per-query optimal F1 (+15.9pp on average) shows that query-level threshold selection could unlock further gains without retraining. Another direction is end-to-end evaluation within a full RAG pipeline, integrating RSRank with upstream embedding retrieval and downstream generation to measure its impact on final answer quality.

\section*{Limitations}

\paragraph{Random negatives vs.\ hard negatives.}
Our BEIR evaluation pairs queries with randomly sampled corpus negatives rather than hard negatives from a retrieval stage. Random negatives could be topically unrelated and easier for any model to distinguish. While this reflects a realistic RAG \emph{context filtering} scenario---where a first-stage retriever has already narrowed the candidate set---it overestimates absolute performance compared to full-corpus retrieval benchmarks. Evaluation with hard negatives from a BM25 or dense retrieval first stage is needed to assess robustness in more adversarial settings.

\paragraph{Per-query calibration.}
While RSRank's zero-threshold design provides better cross-dataset stability than baselines (Sec.~\ref{sec:motivation}), query-level calibration remains imperfect. On HotpotQA, the gap between F1@$\tau$ and F1@$\tau^*$ indicates that many individual queries would benefit from a query-specific threshold. Improving per-query calibration is an important direction for future work.

\section*{Ethical Considerations}
We used AI assistants for support tasks such as improving writing clarity, grammar, and formatting. All technical content, experimental design, analyses, and conclusions were developed, and verified by the authors.

\bibliography{custom}
\newpage
\appendix

\section{Discrete Difference Calculus on Prefix Space}
\label{app:discrete-calculus}

This appendix provides a formal viewpoint for the feature construction in Sec.~\ref{sec:finite-diff-features}. The main paper uses the first-order, length-matched difference features (Equations~\eqref{eq:delta-v}--\eqref{eq:delta-tensor}). The material below justifies terminology such as ``finite difference'' and clarifies how deltas behave under sequences of prefix edits.

\subsection{Shift Operators on Prefixes}

Let $\Sigma$ be the vocabulary and $\Sigma^*$ the set of all finite token sequences. We view $\Sigma^*$ as a rooted directed graph (a $|\Sigma|$-ary tree) whose vertices are prefixes $s\in\Sigma^*$ and whose edges correspond to appending a token: $s \rightarrow s\cdot a$ for any $a\in\Sigma$.

For a function $F:\Sigma^*\rightarrow \mathbb{R}^k$, define the \emph{shift operator}
$S_a$ (append-$a$) by
\begin{equation}
(S_a F)(s) := F(s\cdot a).
\end{equation}
The associated \emph{forward difference} operator is
\begin{equation}
(\Delta_a F)(s) := (S_a F)(s) - F(s) = F(s\cdot a) - F(s).
\label{eq:delta-a}
\end{equation}
Equation~\eqref{eq:delta-a} is the standard discrete/finite-difference construction ``difference = shift minus identity'' in a non-numeric domain.

\subsection{Telescoping Identity}

Finite differences compose along a path in the prefix graph. Let
$u=(u_1,\ldots,u_m)\in\Sigma^m$ be a suffix and $u_{<i}=(u_1,\ldots,u_{i-1})$.
Then for any $F:\Sigma^*\to\mathbb{R}^k$,
\begin{equation}
\begin{split}
F(s\cdot u) - F(s)
&=\sum_{i=1}^{m} \Big( F(s\cdot u_{<i}\cdot u_i) - F(s\cdot u_{<i}) \Big)
\\
&=\sum_{i=1}^{m} (\Delta_{u_i} F)(s\cdot u_{<i})    
\end{split}
\label{eq:app-telescoping}
\end{equation}
Equation~\eqref{eq:app-telescoping} is the discrete analogue of the fundamental theorem of
calculus: the total change equals the sum of incremental changes along a path.

\subsection{Instantiation for Decoder-Only Transformers}

Fix layer $\ell$ and head $h$. For an input sequence $x$, let
$v^{(\ell,h)}_i(x)\in\mathbb{R}^{d_h}$ denote the pre-attention value vector at position $i$.
For a fixed query $q\in\Sigma^m$ and document $d\in\Sigma^n$, define the document-conditioned
value vector for query position $t$:
\begin{equation}
F^{(\ell,h)}_t(d;q) := v^{(\ell,h)}_{n+t}(d\cdot q).
\end{equation}
The main paper's controlled finite difference is exactly the first-order difference
between $d$ and a length-matched null prefix $\varnothing_n$:
\begin{equation}
\delta v^{(\ell,h)}_t(d;q) = F^{(\ell,h)}_t(d;q) - F^{(\ell,h)}_t(\varnothing_n;q),
\end{equation}
and the representational shift tensor $\boldsymbol{\Delta}(d,q)$ collects these deltas across $(\ell,h,t)$.

\section{Ablations} 
\label{app:Ablations}
We conducted ablation studies on our techniques to identify the most effective variations and determine which configurations yield the best results.

\subsection{Representation and Optimization Methods}
\label{sec:shift_ablation}

Here we analyze alternative representation and optimization methods.
Table~\ref{tab:shift_ablation} compares our learned approach on shifts against: (a) learning on raw representations without subtraction to see the effect of ``shifting'', and (b) three closed-form projections on shifts to compare analytical methods against our learned optimisation.

\paragraph{Shift vs.\ raw representations.}
The shift is critical for calibration:while the direct approach achieves comparable R@5(98.3vs97.8), its F1@0 drops by 6.4pp because raw representations lack the centering that makes zero a natural threshold.

\paragraph{Closed-form baselines.}
\sloppy
We consider three analytical solutions:
\begin{enumerate}[leftmargin=1.2em,itemsep=0pt,topsep=1pt,parsep=0pt,partopsep=0pt]
    \item \emph{Oracle alignment:} $\mathbf{B}_{\ell,h} = \mathbf{1}_P \boldsymbol{\mu}_{\ell,h}^\top / \|\boldsymbol{\mu}_{\ell,h}\|^2$, where $\boldsymbol{\mu}$ is the mean shift, so that $\mathbf{B}\boldsymbol{\mu} = \mathbf{1}_P$.
    \item \emph{Separation:} solves a ridge regression $\mathbf{B}^\top = (\mathbf{X}^\top\mathbf{X} + \lambda\mathbf{I})^{-1}\mathbf{X}^\top\mathbf{Y}$ with targets $\mathbf{1}_P$, $\mathbf{0}_P$ for relevant, irrelevant shifts.
    \item \emph{Combined:} adds an oracle constraint to the objective: $\min\|\mathbf{B}\mathbf{S}_{\text{irrel}}\|^2$ subject to $\mathbf{B}\boldsymbol{\mu}{=}\mathbf{1}_P$, yielding a covariance-weighted projection.
\end{enumerate}

All three preserve reasonable ranking (R@5 up to 90.8) but fail at calibration (F1@0 $\leq$ 30.8). These methods optimize alignment and separation targets in projection space, but scoring uses cosine similarity with $\mathbf{1}_P$, which introduces a non-linear normalization. As a result, pushing irrelevant projections toward $\mathbf{0}_P$ does not yield negative scores---it yields near-zero-norm vectors with noisy cosine similarities. Our learned objective optimizes directly on the \emph{scores} (cosine similarities), explicitly pushing relevant scores above zero and irrelevant scores below zero with a margin.

\begin{table}[t]
\centering
\caption{Shift vs.\ direct representations on 2WikiMQA. The shift is essential for threshold calibration (F1@$\tau$).}
\label{tab:shift_ablation}
\small
\begin{tabular}{lcc}
\toprule
\textbf{Method} & \textbf{R@5} & \textbf{F1@$\tau$} \\
\midrule
Learned on shifts (ours)          & 97.8 & \textbf{82.2} \\
Learned on raw repr.\ (no shift)  & \textbf{98.3} & 75.8 \\
\midrule
Closed-form oracle      & 75.3 & 17.9 \\
Closed-form separation  & 90.8 & 23.1 \\
Closed-form combined    & 86.8 & 30.8 \\
\bottomrule
\end{tabular}
\end{table}

\subsection{Choice of Internal Signal}
\label{sec:extraction_ablation}

We compare three internal LLM signals: value vectors (our choice), key vectors, and hidden states. Table~\ref{tab:extraction_ablation} shows that value vectors achieve the best performance, closely followed by key vectors. Hidden states perform worst, suggesting that the head-level decomposition provides useful structure for relevance detection.

\begin{table}[t]
\centering
\caption{Extraction type ablation on 2WikiMQA. Value vectors provide the best signal.}
\label{tab:extraction_ablation}
\small
\begin{tabular}{lcc}
\toprule
\textbf{Signal} & \textbf{R@5} & \textbf{F1@5} \\
\midrule
Value vectors (ours) & \textbf{98.3} & \textbf{66.3} \\
Key vectors          & 97.8 & 66.0 \\
Hidden states        & 96.5 & 65.0 \\
\bottomrule
\end{tabular}
\end{table}

\subsection{Loss Components Ablation}
\label{sec:loss_ablation}

Table~\ref{tab:loss_ablation} shows the effect of removing each loss component. The calibration loss is the most critical: removing it drops F1@$\tau$ by 8.3pp, as the model loses its ability to anchor the decision boundary at zero. Removing the margin loss causes a moderate 1.9-point F1@$\tau$ drop by weakening class separation. Orthogonality regularization and oracle alignment have minimal individual impact.

\begin{table}[htbp]
\centering
\caption{Loss component ablation on 2WikiMQA validation set. $\Delta$ is relative to the full model.}
\label{tab:loss_ablation}
\footnotesize
\setlength{\tabcolsep}{4pt}
\begin{tabular}{lcccc}
\toprule
\textbf{Configuration} & \textbf{R@5} & \textbf{F1@$\tau$} & $\boldsymbol{\Delta}$\textbf{R@5} & $\boldsymbol{\Delta}$\textbf{F1@$\tau$} \\
\midrule
Full model             & 94.0 & 80.2 & ---    & ---    \\
$-$ Calibration loss   & 88.2 & 71.9 & $-$5.8 & $-$8.3 \\
$-$ Margin loss        & 92.9 & 78.3 & $-$1.1 & $-$1.9 \\
$-$ Ortho.\ reg.       & 93.4 & 79.9 & $-$0.6 & $-$0.3 \\
$-$ Oracle alignment   & 93.3 & 80.0 & $-$0.7 & $-$0.2 \\
$-$ Norm constraint    & 93.6 & 79.4 & $-$0.4 & $-$0.8 \\
\bottomrule
\end{tabular}
\end{table}
\newpage
\subsection{Training Details}
\label{app:training-details}

\begin{table}[htbp]
\centering
\small
\caption{Training hyperparameters used across all experiments.}
\begin{tabular}{lc}
\toprule
\textbf{Hyperparameter} & \textbf{Value} \\
\midrule
Optimizer & AdamW \\
Learning rate schedule & Cosine annealing \\
Initial learning rate & $0.05$ \\
Final learning rate & $0.001$ \\
Weight decay & $10^{-3}$ \\
Training precision & Mixed precision \\
Maximum epochs & $120$ \\
Early stopping patience & $35$ \\
$\lambda_{\text{cal}}$ & $0.8$ \\
$\lambda_{\text{align}}$ & $0.5$ \\
$\lambda_{\text{ortho}}$ & $0.05$ \\
$\lambda_{\text{frob}}$ & $0.1$ \\
Margin $(m)$ & $0.5$ \\
Projection dimension $(P)$ & $64$ \\
\bottomrule
\end{tabular}
\label{tab:training-hparams}
\end{table}
\end{document}